\documentclass[onecolumn,showpacs,showkeys,amsmath,nofootinbib,amssymb,superscriptaddress]{revtex4}
\usepackage[dvips]{color}
\usepackage{epsfig}
\usepackage{amsmath}
\usepackage{graphicx}

\begin{document}

\title{Testing Quantum Gravity through Dumb Holes}

\author{Behnam Pourhassan}\email{ b.pourhassan@du.ac.ir}
\affiliation{School of Physics, Damghan University, Damghan, Iran}

\author{Mir Faizal }\email{f2mir@uwaterloo.ca}
\affiliation{Irving K. Barber School of Arts and Sciences,  University of British Columbia - Okanagan, Kelowna,   BC V1V 1V7, Canada}
\affiliation{Department of Physics and Astronomy, University of Lethbridge,  Lethbridge, AB T1K 3M4, Canada}

\author{Salvatore Capozziello}\email{capozzie@na.infn.it}
\affiliation{Dipartimento di Fisica, Universit\`{a} di Napoli "Frederico II" Complesso Universitario di Monte S. Angelo, Edificio G, Via Cinthia, I-80126 Napoli, Italy}
\affiliation{Gran Sasso Science Institute (INFN), Via F. Crispi 7, I-67100, L' Aquila, Italy}

\date{\today}

\begin{abstract}
We propose a method to test the effects of quantum fluctuations  on  black holes by
analyzing the effects of thermal fluctuations on dumb holes, the analogues for black holes. The proposal is based on the Jacobson formalism,
where the  Einstein field equations are viewed as  thermodynamical relations, and so the
quantum  fluctuations   are generated  from the  thermal fluctuations.
It is well known that all  approaches to quantum gravity generate  logarithmic corrections to the entropy of a black hole
and  the coefficient of this term varies according to the   different   approaches to the quantum gravity.
It is possible to  demonstrate that such logarithmic terms are also generated from  thermal fluctuations in dumb holes.
In this paper, we claim that it is possible to experimentally test such corrections for dumb holes, and also obtain the correct coefficient for them.
This fact can then be used to predict the effects of quantum fluctuations on realistic  black holes, and so it  can also be used, in principle,  to
experimentally test  the different approaches to quantum gravity.
\end{abstract}

\maketitle

%\section{Introduction}

The entropy of a black hole  ($s$) scales with its area rather than its volume, $s = A/4$, where $A$ is the area of its horizon \cite{1, 1a}. This is true not only in general relativity but also in modified theories of
 gravity, and so even in modified theories of gravity the leading order  entropy scales with the area
 \cite{addazi,report,teleparallel, vb12, vb14}.
 This expression for the entropy of a black hole can be obtained
 using a semi-classical approximation. Since any approach to quantum gravity has to agree with the semi-classical approximation at a sufficiently
 low energy scale, it is expected that all the approaches to quantum gravity will produce this expression for the
 entropy of a black hole. In fact, it has been observed that
 even though there are various different approaches to
quantum gravity, all of them predict that the entropy of a black hole would be related to the area of the horizon as
$s = A/4$. However, all these
different approaches to quantum gravity also predict corrections to this relation between the area and the entropy.
Non-perturbative quantum general relativity  has been used
to obtain such a logarithmic correction to the area-entropy law   \cite{1z}.  It is known that
  a   relation exists between the
  density of states of a black hole and
the conformal blocks of a    conformal field theory  in
non-perturbative quantum general relativity. This relation is used to obtain the
logarithmic correction to the area-entropy law. It has also been demonstrated that such a logarithmic
term can be obtained using the Cardy formula for the entropy of a black hole \cite{card}.
It is also possible to analyze the exact partition function for a Ba$\tilde{\mbox{n}}$ados-Teitelboim-Zanelli (BTZ) black hole \cite{2a}, and calculate the correction to the area-entropy law for such a  black hole.
This calculation points out  that  the entropy of a BTZ black hole results corrected by a logarithmic term \cite{gks}. It has also been observed that the entropy of a dilatonic black holes   is also  corrected by a logarithmic term \cite{jy}. The string theory corrections  to the area-entropy law of a back hole have been obtained, and it has been observed that such corrections are logarithmic \cite{solo1, solo2, solo4, solo5}. The correction term obtained  from  the generalized uncertainty principle is also a logarithmic term \cite{mi, r1}.

In the recent work  \cite{1605.03458}, the logarithmic correction applied to the dyonic charged anti-de Sitter black hole, and holographic dual Van der waals fluid has been studied. Another kind of logarithmic correction already considered for the G\"{o}del black hole \cite{PJFP}. Many recently, we study logarithmic correction of a sufficient small singly spinning Kerr-AdS black hole \cite{NPB}.

It may be noted that the Wald entropy produced by the gravity containing higher curvature terms
coincides with the entropy corrections produced by generalized uncertainty principle \cite{facd12}.  So, we expect the modified
theories of gravity with higher curvature terms would also produce such corrections. This might be because these higher
curvature terms can be viewed as higher loop quantum corrections using the formalism of the effective field theories \cite{efcd12}.
Such corrections have also been obtained using the Cardy-Verlinde formula \cite{c1a2,c2a1}.
Thus, the  quantum corrections from these various approaches to quantum gravity produce similar scaling behavior
of the leading order corrections to the entropy.

Now, it is possible to state that all the different approaches to quantum gravity generate  logarithmic corrections to the
area-entropy law of a black hole.
It is worth  noticing that even   though the leading order corrections to this area-entropy law are logarithmic corrections of the form $\ln A$, the
coefficient of such a term depends on  the  approach to the quantum gravity. The coefficient
of this term can be used  as a parameter to differentiate among different approaches to the quantum gravity.
Since the values of the coefficients of the logarithmic terms depend on the  chosen approach to quantum gravity, it can be
argued that such terms are generated from quantum fluctuations of the  space-time geometry rather than from  matter fields on
that space-time. However, such  quantum fluctuations only become  important at  very small scales. So, these
quantum fluctuations start  to become important as the size of the black hole reduces
  and its temperature increases. Even though it is possible to neglect the effect of quantum fluctuations for very large black holes, it
  is not possible to neglect such corrections for  sufficiently small  black hole with very
high  temperature.

Thus, the temperature of a black hole can be used as an indicator of the scale at which
quantum fluctuations become important. This is obvious in   the Jacobson formalism where the
   Einstein field equations are viewed as thermodynamical relations \cite{z12j, jz12}. In fact, in the Jacobson formalism,  the Einstein  equations are
   obtained by requiring the
Clausius relation  holding   at the local Rindler causal horizons through each space-time point \cite{z12j, jz12}.

Now if the  Einstein equations are thermodynamical  relations, then
   quantum fluctuations in the geometry of the space-time are generated from the   thermal fluctuations in the thermodynamical
   system used to derive the Einstein field equations. As soon as these thermal fluctuations become relevant at high temperature,
   and high temperature corresponds to the small scale for black holes, the quantum fluctuations in the metric of a black hole
 can also be analyzed using thermal fluctuations in  thermodynamics.

 It has been demonstrated that the
  thermal fluctuations  in the thermodynamics of a black hole  correct the area-entropy law by a logarithmic correction term \cite{l1, SPR}.
 In fact, the area-entropy law results modified due to thermal fluctuations as
$S=s - \ln{s_1}/ 2,$ where $s= A/4$ and  $s_1 \sim s T^{2}$. Now as this   logarithmic term occurs in  all approaches to quantum gravity,
but its coefficient depends on the specific approach to quantum gravity, we will write this entropy correction as
$S=s - a \ln{s T^2}/ 2$, where we have added an arbitrary parameter $a$. The exact value of $a$ can be used to discriminate among
different  quantum gravity models. Such an approach has been used for analyzing the effects of
thermal fluctuations on the  thermodynamics of an AdS black holes  \cite{adbc},
a black Saturn \cite{dabc}, charged  dilatonic black Saturn \cite{dabc1}, and a modified Hayward black hole \cite{dabc2}.

Even though this seems to be an interesting way to test models of quantum gravity using black hole thermodynamics, the problem with this approach
is that it is almost impossible, from an experimental point of view,  to test the effect of quantum fluctuations for real black holes. However, it is possible to test such effects for
dumb holes, which are analogous black hole as  solutions  \cite{1011.3822,dansj,dcnbj,d1}.

In fact, it is well known that inhomogeneous fluid flows can become supersonic and   can   produce acoustic
analogs of   black holes.  Such holes are called \textit{dumb holes}, and they can have a
 Hawking-like radiation of phonons
with a temperature  given by the gradient of the velocity at the horizon \cite{d}.
Thus, one can analyze, in principle,  the effects that thermal fluctuations have on dumb holes. However, it is also possible to experimentally test the behavior
of dumb holes \cite{expe1}, and so one can
  compare the analysis to the experimental data. In summary, we can experimentally verify
the effects of thermal fluctuations on the thermodynamics of dumb holes.

However, since dumb holes are, in general,  mathematically equivalent to the black holes, and, in the Jacobson formalism,  even the real black holes are thermodynamical
objects, one can use the analysis performed  on dumb holes to fix the value of $a$ for black holes.   Practically,  by calculating
and observing the value of $a$ from dumb holes, one  can infer the correct model  of quantum gravity since the coefficient of the
logarithmic term strictly depends on the approach to the quantum gravity.

In this letter we want to investigate
the effects of thermal fluctuations on the thermodynamics of dumb holes in order to put constraints on the value of the coefficient of the logarithmic correction.

The construction of dumb holes is well known and its gravity dual has also been analyzed  in detail \cite{1011.3822}.
The metric of the dumb hole can be written as
\begin{equation}\label{1}
ds^{2}=\sqrt{3}\mathcal{T}\left[-(1-\frac{2}{3}\gamma(z)^{2})d\tau^{2}+\frac{dz^{2}}{3(1-\frac{2}{3}\gamma(z)^{2})}+R(z)^{2}d\Omega_{2}^{2}\right],
\end{equation}
where $\mathcal{T}$ is a quantity with  energy dimensions. It
is  usually assumed  to be a function of the temperature $T$ and the chemical potential $\mu$ (or charge $q$).
For the uncharged case ($\mu=0$ or $q=0$), we have $\mathcal{T}=T$. It can also be expressed in terms of a scalar field $\phi$, as
${\mathcal{T}}^{2}=-(\partial_{\mu}\phi)(\partial^{\mu}\phi)$. Here $\gamma(z)$ is a Lorentz factor defined in terms of the velocity $v(z)$,
\begin{equation}\label{2}
\gamma(z)=\frac{1}{1-v(z)^{2}}.
\end{equation}
The curvature corresponding to this metric can be denoted as    $R(z)$. It is worth noticing  that    $R(z)=z$ corresponds to a flat geometry.
 Now the expression for  the acoustic Hawking temperature $T_{H}$ can be written as
\begin{equation}\label{3}
T_{H}=\frac{3}{4\pi}\left(\frac{dv_{z}}{dz}\right)_{z=z_{h}}.
\end{equation}
This temperature is obtained  by using the  standard techniques of  Euclidean  quantum gravity
near the horizon $z_{h}$. The only non-zero component of the velocity  is $v_{z}$, and then \cite{1011.3822}
\begin{equation}\label{4}
v_{z}-v_{z}^{3}=\frac{\Phi_{s}}{\mathcal{T}_{\infty}^{3}}R(z)^{-2},
\end{equation}
where $\mathcal{T}_{\infty}$ and $\Phi_{s}$ denote asymptotic temperature and entropy,  respectively.
The  gravitational dual  of this model
has been analyzed using the    fluid-gravity correspondence \cite{1107.5780}.
The gravity dual action is obtained from type IIB supergravity with negative cosmological constant $\Lambda=-6$ \cite{0809.2596},
\begin{equation}\label{5}
S=\frac{1}{16\pi G}\int{d^{5}x\left[R+12-F_{AB}F^{AB}-\frac{4\kappa}{3}\varepsilon^{EABCD}A_{E}F_{AB}F_{CD}\right]},
\end{equation}
with the Chern-Simons parameter $\kappa= 1/(2\sqrt{3})$, and $G$ is the Newton constant in five dimension. Special case of $\kappa= 0$ corresponds to a pure Maxwell theory with no Chern-Simons type interactions. Also $A_{E}$ ($E=0,1,2,3,4$) is a $U(1)$ gauge field with strength $F_{AB}$.
So, the resulting charged black brane solution is  given by,
\begin{eqnarray}\label{6}
ds^{2}&=&-2u_{\mu}dx^{\mu}dr-r^{2}f(r)u_{\mu}u_{\nu}dx^{\mu}dx^{\nu}+r^{2}P_{\mu\nu}dx^{\mu}dx^{\nu},\nonumber\\
A&=&\frac{\sqrt{3}Q}{2r^{2}}u_{\mu}dx^{\mu},\nonumber\\
f(r)&=&1-\frac{m}{r^{4}}+\frac{Q^{2}}{r^{6}},
\end{eqnarray}
where $m$ and $Q$ as mass and charge parameters, and they  are related to the fluid parameters by the  gauge-gravity dictionary \cite{0809.2596},
\begin{eqnarray}\label{7}
\epsilon&=&3\alpha m,\nonumber\\
q&=&\sqrt{3}\alpha Q,
\end{eqnarray}
where $\alpha\equiv(16\pi G)^{-1}$.
From  the fluid side of this duality,  the first law of thermodynamics can be expressed as
\begin{equation}\label{8}
d\epsilon=Tds+\mu dq,
\end{equation}
where $s$ denotes uncorrected entropy,
and $\epsilon$ is the energy density which is related to the pressure via $\epsilon=3p$. Now, by using the fluid-gravity correspondence, it can be
demonstrated  that \cite{1011.3822}
\begin{eqnarray}\label{9}
\epsilon&=&3\alpha \left(\frac{s}{4\pi\alpha}\right)^{\frac{4}{3}}+\frac{q^{2}}{\alpha}\left(\frac{s}{4\pi\alpha}\right)^{-\frac{2}{3}},\nonumber\\
T&=&\frac{1}{\pi} \left(\frac{s}{4\pi\alpha}\right)^{\frac{1}{3}}\left(1-\frac{8\pi^{2}q^{2}}{3s^{2}}\right),\nonumber\\
\mu&=&\frac{2q}{\alpha}\left(\frac{s}{4\pi\alpha}\right)^{-\frac{2}{3}}.
\end{eqnarray}
The effect of thermal fluctuations for this system can be analyzed  by
relating it to a conformal field theory \cite{l1},  and
using the modular invariance of the partition function of the conformal field theory \cite{card}.
Now if  $S= A\beta_{\kappa}^l + B \beta_{\kappa}^{-j} $,
where $\beta_{\kappa}=(\kappa_{B}T)^{-1}$, and  $l, j, A, B > 0$, then the relation between the
  corrected entropy  $S$ and the  original entropy $s$, can be written as \cite{l1, dabc1}
\begin{equation}\label{14}
S = s -\frac{a}{2} \ln |sT^{2}|+\cdots,
\end{equation}
where the  variable $a$ is again introduced to parameterize the effect of thermal fluctuations on the thermodynamics
of dumb hole.
The original entropy and partition function are related to each other by the following differential equation,
\begin{equation}\label{15}
s=X+T\frac{dX}{dT}.
\end{equation}
Here $X\equiv \kappa_{B}\ln{Z}$ has been introduced to simplify relations, where  $Z$ is the partition function.
Now we can obtain partition function from
\begin{equation}\label{16}
X=\frac{1}{T}\left[\int{s dT}+c\right],
\end{equation}
where $c$ is an integration constant. Thus,  the partition function can be obtained if
we know the temperature dependence of the  entropy.
In the case of  entropy corrected by thermal fluctuation,
we can use Eqs. (\ref{14}) and (\ref{15}) to obtain the modified partition function,
\begin{equation}\label{17}
X=\frac{1}{T}\left[\int{g(S) dT}+C\right],
\end{equation}
where $C$ is another integration constant and,
\begin{equation}\label{18}
g(S)=-\frac{a}{2}LW\left(-\frac{2e^{-\frac{2S}{a}}}{aT^{2}}\right),
\end{equation}
where $g(S)$ is obtained from Eq. (\ref{14}) as the Lambert  $W$ function $LW$. In the case of $a=0$ where $g(S)=s$ we have ordinary thermodynamics.
The  partition function given by Eq. (\ref{17}) can be used    to obtain   thermodynamics information about this system.
Thus, we write  internal energy as
\begin{equation}\label{19}
U=T^{2}\frac{dX}{dT},
\end{equation}
and then specific heat,
\begin{equation}\label{20}
C_{v}=\frac{dU}{dT}.
\end{equation}
Obtaining  general
analytical solutions for Eq. (\ref{17}) is very difficult.
In this paper,   we  restrict the analysis  to uncharged dumb holes ($q=0$).  In this case, it
  is possible to obtain analytical results. Now,
    $s=4\alpha\pi^{4}T^{3}$ is obtained using Eqs. (\ref{9}). Therefore,   entropy corrected by thermal fluctuations is given by,
\begin{equation}\label{21}
S=4\alpha\pi^{4}T^{3}-\frac{a}{2}\ln{4\alpha\pi^{4}T^{3}}.
\end{equation}
The function $g(S)$ is given by  Eq. (\ref{18}), and it goes to infinity at zero temperature.
For  finite temperature in the range  $0<T<\infty$,  it behaves as follow,
\begin{equation}\label{22}
g(S)=\frac{e^{-\frac{2S}{a}}}{T^{2}}+\mathcal{O}\left(\frac{1}{T^{4}}\right).
\end{equation}
However, in the  zero-temperature limit $T\rightarrow0$,  we have  $\frac{1}{T^{2}}\rightarrow\infty$. The
 thermal fluctuations vanishes for $a=0$. Furthermore,
 in  the limit  $T\rightarrow\infty$, it is $g\rightarrow \mbox{constant}$,  with infinitesimal values.
 Hence $\kappa_{B}\ln{Z}\approx c_{1}+\frac{c_{2}}{T}$. Now using  Eqs. (\ref{21}) and (\ref{22}),  we obtain
\begin{equation}\label{23}
g(S)\approx 4\alpha\pi^{4}T^{3}e^{-\frac{8\alpha}{a}\pi^{4}T^{3}}.
\end{equation}
Therefore, by using  Eqs.  (\ref{23}) in (\ref{17}),  we also obtain
\begin{equation}\label{24}
X=AT^{-\frac{1}{2}}e^{-\frac{4\alpha\pi^{4}}{a}T^{3}}WM\left(\frac{1}{6},\frac{2}{3},\frac{8\alpha\pi^{4}}{a}T^{3}\right)+\frac{C}{T},
\end{equation}
where $A\equiv\frac{\sqrt{2}a^{\frac{7}{6}}}{16\pi^{\frac{2}{3}}\alpha^{\frac{1}{6}}}$,
and $WM\left(\frac{1}{6},\frac{2}{3},h\right)$ is Whittaker function $M$
with $h\equiv\frac{8\alpha\pi^{4}}{a}T^{3}$. Before analyzing  the general solution, using expression (\ref{24}), we will use approximations for
 some   special cases.
At the first order approximation, we obtain,
\begin{equation}\label{25}
Z=e^{B_{1}T^{3}+\frac{C}{T}},
\end{equation}
where $B_{1}$ is a constant which depends on $\alpha$. Thus, we get
\begin{eqnarray}\label{26}
U&=&3B_{1}T^{4}-C,\nonumber\\
C_{v}&=&12B_{1}T^{3},
\end{eqnarray}
which are derived from the original entropy  given in Eq. (\ref{16}).
Now the effects of thermal fluctuations can be obtained by using  higher order corrections
to the Whittaker function $M$  given  in  Eq. (\ref{24}). At the second order approximation, we obtain
\begin{equation}\label{27}
Z=e^{B_{1}T^{3}+B_{2}T^{6}+\frac{C}{T}},
\end{equation}
where $B_{1}$ and $B{2}$ are  constants  depending on $\alpha$. We can write
\begin{eqnarray}\label{28}
U&=&3B_{1}T^{4}+6B_{2}T^{7}-C,\nonumber\\
C_{v}&=&12B_{1}T^{3}+42B_{2}T^{6}.
\end{eqnarray}
Hence, we obtained the effect of thermal fluctuations parameterized by the  coefficients $B_{1,2}$.
It is clear that, by the increasing temperature, these thermal fluctuations  become  important, while,
  at lower temperatures, they can be neglected. These are only approximate solutions.  We need now  the  general case.
It is  given by Eq. (\ref{24}).  We can plot $U$ and $C_{v}$ in terms of $T^{3}$ (see Fig. \ref{1}).

\begin{figure}[h!]
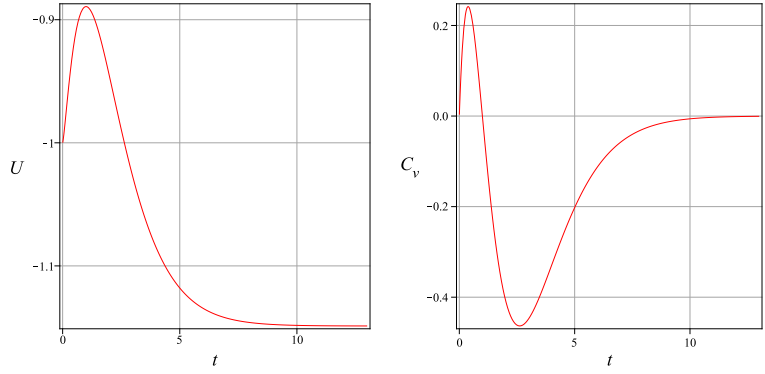

 \begin{center}$
 \begin{array}{cccc}
\includegraphics[width=50 mm]{1-1.eps}&\includegraphics[width=50 mm]{1-2.eps}
 \end{array}$
 \end{center}
\caption{Internal energy and Specific heat in terms of $t\equiv T^{3}$ with $C=a=1$ and $G=\frac{\pi^{3}}{2}$.}
 \label{1}
\end{figure}

It is easy to  see that the  internal energy decreases at infinitesimal temperature where thermal fluctuations can be neglected.
However, if these  thermal fluctuations are not neglected, then  the  internal energy increases till a
critical temperature $T_{c}$.
Thus, the thermal fluctuations  occur at  $T_{min}\leq T\leq T_{c}$.
As we can see from the second plot, system will become unstable at $T\geq T_{c}$,
and   we obtain   $C_{v}=0$ at very high temperature.
\vspace{3.mm}

In this letter, we have  analyzed  the   effect of   thermal fluctuations  on the thermodynamics  of
 acoustic dumb holes  occurring in  inhomogeneous supersonic fluids.
 It was   observed that these thermal fluctuations correct the entropy of a dumb hole by  a
  logarithmic term.  We then argued that since it is well
  known that all approaches to quantum gravity generate  logarithmic corrections to the entropy of a black hole,
and  the coefficient of this term depends on  the quantum gravity approach, the coefficient of this logarithmic can be
used as a tool to test the correct quantum  gravity model. It is important to stress  that
 the Einstein equations can be viewed as  thermodynamical relations in the Jacobson formalism, and so
 the quantum  fluctuations   can be obtained from
 thermal fluctuations in this formalism. Here, we proposed that
  the effect of thermal fluctuations on the thermodynamics of dumb holes can be used to predict the behavior of quantum fluctuations on
real black holes in the Jacobson formalism. These analog effects could be experimentally measured for a dumb hole,
and thus we can experimentally obtain, in principle,   the exact value of the coefficient of  logarithmic correction for dumb holes.
 Thus, we can expect to predict the effect of thermal fluctuations for real black holes, if we know
its behavior of  dumb holes. However, as the coefficient of the logarithmic term depends on the approach to quantum gravity, we can
also use this analysis, in general,  to test the correct approach to quantum gravity.

For the future works it is interesting to apply our method to some important kinds of black holes like a new regular black hole \cite{new reg}, Horava-Lifshitz black hole \cite{HL}, Myers-Perry black hole \cite{MP} or STU black hole \cite{JHEP, STU}. It would also be interesting to analyze the effect of
logarithmic term on the holographic heat engines \cite{engine} dual to the mentioned black holes. Finally it ie interesting to consider charged dumb hole and study the effect of thermal fluctuations on the charged dumb hole.

\end{document}